\documentclass[12pt,preprint]{aastex}

\shorttitle{}
\shortauthors{Nesvorn\'y et al.}

\begin{document}
\baselineskip 19.pt

\title{Bi-lobed Shape of Comet 67P from a Collapsed Binary}

\author{David Nesvorn\'y$^1$, Joel Parker$^1$, David Vokrouhlick\'y$^{1,2}$}
\affil{(1) Department of Space Studies, Southwest Research Institute,\\
1050 Walnut St., Suite 300, Boulder, CO, 80302, USA}
\affil{(2) Institute of Astronomy, Charles University, \\
V Hole\v{s}ovi\v{c}k\'ach 2, CZ--18000 Prague 8, Czech Republic}

\begin{abstract}
The Rosetta spacecraft observations revealed that the nucleus of comet 67P/Churyumov-Gerasimenko
consists of two similarly sized lobes connected by a narrow neck. Here we evaluate the possibility
that 67P is a collapsed binary. We assume that the progenitor of 67P was a binary and consider
various physical mechanisms that could have brought the binary components together, including
small-scale impacts and gravitational encounters with planets. We find that 67P could be a primordial 
body (i.e., not a collisional fragment) if the outer planetesimal disk lasted $\lesssim$10 Myr before 
it was dispersed by migrating Neptune. The probability of binary collapse by impact is $\simeq$30\% 
for tightly bound binaries. Most km-class binaries become collisionally dissolved. Roughly 10\% of 
the surviving binaries later evolve to become contact binaries during the disk dispersal, when bodies 
suffer gravitational encounters to Neptune. Overall, the processes described in this work do not seem 
to be efficient enough to explain the large fraction ($\sim$67\%) of bi-lobed cometary nuclei 
inferred from spacecraft imaging. 
\end{abstract}


\section{Introduction}

\begin{figure}
\epsscale{0.8}
\plotone{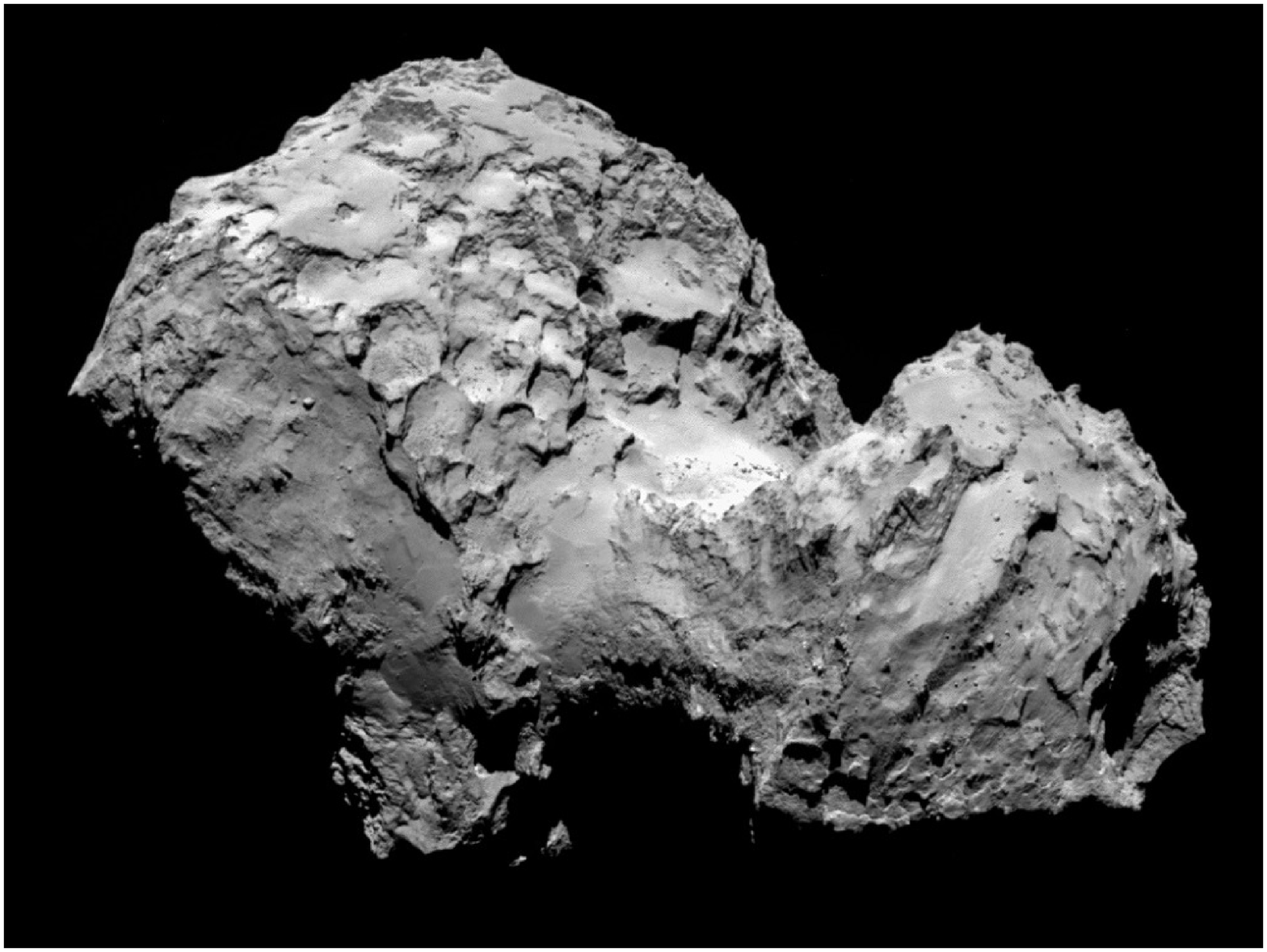}
\caption{Comet 67P/Churyumov-Gerasimenko as imaged by the OSIRIS camera onboard of the Rosetta spacecraft 
on August 3, 2014. Credit: ESA/Rosetta/MPS and the OSIRIS team.}
\label{67P}
\end{figure}

The spectacular images of Rosetta's OSIRIS camera revealed that the nucleus of comet 67P/Churyumov-Gerasimenko 
has a bi-lobed shape (Figure \ref{67P}). The dimensions of the small and large lobes are 
$2.5\times2.1\times1.6$ km and $4.1\times3.5\times1.6$ km, respectively (Jorda et al. 2016). 
Their volume ratio is 2.4. The two lobes are connected by a narrow neck giving 67P appearance 
of a contact binary. Several recent studies addressed the question of the origin of the 67P shape. 
Jutzi \& Benz (2017) suggested that 67P formed as a result of the low-energy, sub-catastrophic 
impact on an elongated, rotating parent body. Schwartz et al. (2018) proposed, instead, that 67P and 
other bilobate comets formed during catastrophic collisional disruptions of much lager bodies. 
Another possibility, which we investigate in this work, is that 67P formed as a binary planetesimal 
that subsequently collapsed to become a contact binary (Rickman et al. 2015). 

Opinions differ about whether 67P is primordial (i.e., formed as a small cometesimal by some accretion 
process 4.6 Gyr ago) or whether it was once part of a larger parent planetesimal
and was liberated from it by an energetic impact. Morbidelli \& Rickman (2015, hereafter MR15) 
studied the collisional survival of 67P in a massive outer disk at 15-30 au, which is thought to be 
the ultimate source of comets (Nesvorn\'y et al. 2017). They found that an object of the 
size of 67P should suffer tens of catastrophic collisions over the assumed disk lifetime (400 Myr). 
The survival probability would be negligible in this case ($<10^{-4}$). The disk lifetime, however, 
may have been shorter (Jutzi et al. 2017 and discussion in Section 2). The survival of 67P after the outer 
disk dispersal, during the stage when 67P spent $>$4 Gyr in the scattered disk, is less of an issue. 

Davidsson et al. (2016), on the other hand, argued that 67P is a primordial rubble pile that 
somehow, perhaps because the outer disk remained dynamically cold, avoided being shattered by 
impacts. They pointed out that the surfaces of both lobes are characterized by thick layers that 
envelope the lobes individually (Massiorini et al. 2015). If the layering was produced during 
the accretion of 67P, then its existence indicates that 67P somehow avoided being collisionally
disrupted, in contradiction to the conclusions of MR15. This would suggest, among other things, 
that the final stage of the 67P nucleus formation was a gentle merger between two similarly-sized 
cometesimals (Rickman et al. 2015).

Blum et al. (2017) took the arguments of Davidsson et al. a step further, proposing that 67P
formed via a gravitational collapse of a bound clump of pebbles. The gravitational collapse,
which can be triggered by the streaming instability in a protoplanetary disk (Youdin \& Goodman 
2005), is a model for the formation of planetesimals that gained substantial support in the
recent years (e.g., Youdin \& Johansen 2007, Johansen et al. 2009, 2012, Nesvorn\'y et al. 2010,
Simon et al. 2017). Blum et al. (2017) pointed out that this formation model is compatible 
with several properties of 67P, including the global porosity, homogeneity, tensile strength, 
thermal inertia, and sizes and porosities of the emitted dust particles. 
 
Here we evaluate the possibility that the bi-lobed shape of 67P emerged in two steps: (1) 67P's 
parent binary formed by the gravitational collapse of pebbles (Nesvorn\'y et al. 2010) or by some 
other mechanism (e.g., Goldreich et al. 2002), and (2) the parent binary was destabilized leading 
to a gentle collision between the binary components. 

As for (1), the binaries with similarly sized components are common in the dynamically cold population 
of the classical Kuiper belt at 40-50 au (the so-called Cold Classicals or CCs; Noll et al. 2008). 
Here we assume that analogous binaries formed in the massive planetesimal disk at 15-30~au. This is 
reasonable because the planetesimal accretion in the outer solar system should have been controlled 
by the same processes (see, e.g., Youdin \& Kenyon 2013 for a review).\footnote{Note that 
loosely-bound binaries do not survive gravitational encounters with Neptune during 
the dynamical implantation of bodies into the Kuiper belt (Parker \& Kavelaars 2010). That is 
why only a very few wide, equal-size binaries were detected in the dynamically hot population of the 
Kuiper belt.} The volume ratio of 67P's lobes indicates $R_2/R_1\simeq0.75$, where $R_1$ and $R_2$ are 
the effective radii of the two lobes. This turns out to be a common value among the known CC binaries 
and is also found to be right in the middle of the values expected from the gravitational collapse 
(Nesvorn\'y et al. 2010). 

As for (2), the parent binary of 67P could have been destabilized by small impacts that transferred  
the linear momentum of the projectile to binary components. Alternatively, it could have been 
destabilized dynamically by gravitational encounters with planets (during the implantation of cometesimals 
into cometary reservoirs or later transfer of bodies from the cometary reservoirs into the inner solar 
system), or by the Kozai cycles (for a binary orbit that had a significant tilt relative to the 
heliocentric orbit). While many destabilized binaries become unbound, some may end up, after low-speed 
collisions between components, as contact binaries. 

Another possibility is that the bi-lobed shape of 67P formed during the gravitational collapse itself. 
During the collapse, the pebble cloud fragments and forms bodies of different sizes (Nesvorn\'y et. al. 
2010). They remain gravitationally bound and collide between themselves at the characteristic speeds 
of $\simeq1$-10 m s$^{-1}$, often resulting in accretion. It is therefore possible that the bi-lobed 
shape of 67P formed as result of such an early merger of two km-sized cometesimals that formed within 
the collapsing cloud. It is not clear, however, whether this process would lead to the formation of 
contact binaries, or whether the accreted agglomerates would collapse into a more spherical object.

The main difference between this alternative and the two-step process discussed above is the time interval 
elapsed between: (i) the formation of components, and (ii) their presumed assembly into a contact binary. 
In the two-step model, there is a significant delay between processes (i) and (ii), possibly as long as 
$\gtrsim$10 Myr, during which the two components of a binary can gain the internal strength (e.g., by modest 
radioactive heating), and resist disintegration during their later assembly into a contact binary.   

\section{Collisional and dynamical history of 67P} 

67P is a Jupiter-family comet (JFC) with the semimajor axis $a=3.46$ au, perihelion distance
$q=1.24$ au, eccentricity $e=0.64$, inclination $i=7.0^\circ$, and orbital period $P=6.4$~yr. Its 
present orbit is unremarkable among JFCs. Several attempts have been made to reconstruct the past 
dynamical history of 67P. For example, Maquet (2015) numerically integrated the orbit of 67P backward
in time. Their integration included the gravitational perturbation of planets, non-gravitational 
forces resulting from 67P's activity, and relativistic effects. They found that 67P suffered 
a close encounter with Jupiter on February 4, 1959, during which the perihelion distance
dropped from $\simeq$2.7 au before the encounter to 1.3 au after the encounter. Another
encounter of 67P to Jupiter occurred on October 2, 1923. 

Reconstructing the orbital history of 67P much further is difficult because 
of orbital chaos. At some point, after several Lyapunov times\footnote{The Lyapunov time
expresses the characteristic timescale for the exponential divergence of nearby orbits.
The motion is generally unpredictable on a timescale of several Lyapunov times.} elapse 
--typically centuries for JFCs --, the backward integration behaves much like a forward 
integration, with the overwhelming majority of orbital clones being ejected from the solar system 
by Jupiter (Guzzo \& Lega 2017). This does not mean, however, that 67P was
injected directly onto Jupiter-crossing orbit from some distant reservoir. Instead, the 
long-term integration tells us only about the {\it future} evolution of 67P (assuming that 
it physically survives that long). Therefore, to establish the past dynamical history of 67P, a 
different approach is needed.

Here we use the JFC model developed in Nesvorn\'y et al. (2017, hereafter N17). N17 performed 
full-scale simulations, in which cometary reservoirs were populated in the early solar system 
and evolved over 4.5 Gyr (see Section 3 for more details). The population of present-day comets 
obtained in the model was compared with the number and orbits of the known JFCs, demonstrating 
good fidelity of the model. The physical lifetime of model JFCs was parametrized in N17 
by the number of perihelion passages below 2.5 au, $N_{\rm P}(2.5)$, and was constrained 
from the comparison with the known population of active JFCs (see also Levison \& Duncan 1997). 
The best fit was found to be dependent on the nucleus size. For 67P (effective radius $R\simeq1.6$ km), 
this work suggests $N_{\rm P}(2.5) \sim 1000$, indicating the physical lifetime on an orbit with 
$q<2.5$ au of $\sim$ 5,000-10,000 years. This is roughly consistent with the measured mass loss 
of 67P, $\sim1.8\times10^{10}$ kg per orbit (Paetzold et al. 2016). Assuming that this represents 
the average activity of 67P and that the mass loss is driven by surface processes, the current 
erosion rate of $\sim$1 meter per orbit indicates that 67P should last $\sim$10,000 yrs. 

We selected model JFCs from N17 that reached orbits similar to that of 67P during 
their dynamical evolution. In practice, the following selection criteria were used: $3.3<a<3.6$~au, 
$0.5<e<0.8$ and $5<i<9^\circ$. For each model comet, we followed its evolution 
from the source reservoir (mainly the scattered disk at $50<a<150$ au) to the time when the selection 
was made. We monitored the number of perihelion passages below 2.5 au and the time spent on a JFC 
orbit after first reaching $q<2.5$ au. If the number of perihelion passages for an individual 
comet exceeded 1000 (i.e., $N_{\rm P}(2.5)=1000$ using the N17 definition of the physical lifetime), 
the comet was not considered (assuming that it would cease to be active, either become dormant 
or disrupt, before it reached the selection time). All other cases were considered together 
to give us a statistical information about the past evolution history of 67P.

\begin{figure}
\epsscale{0.8}
\plotone{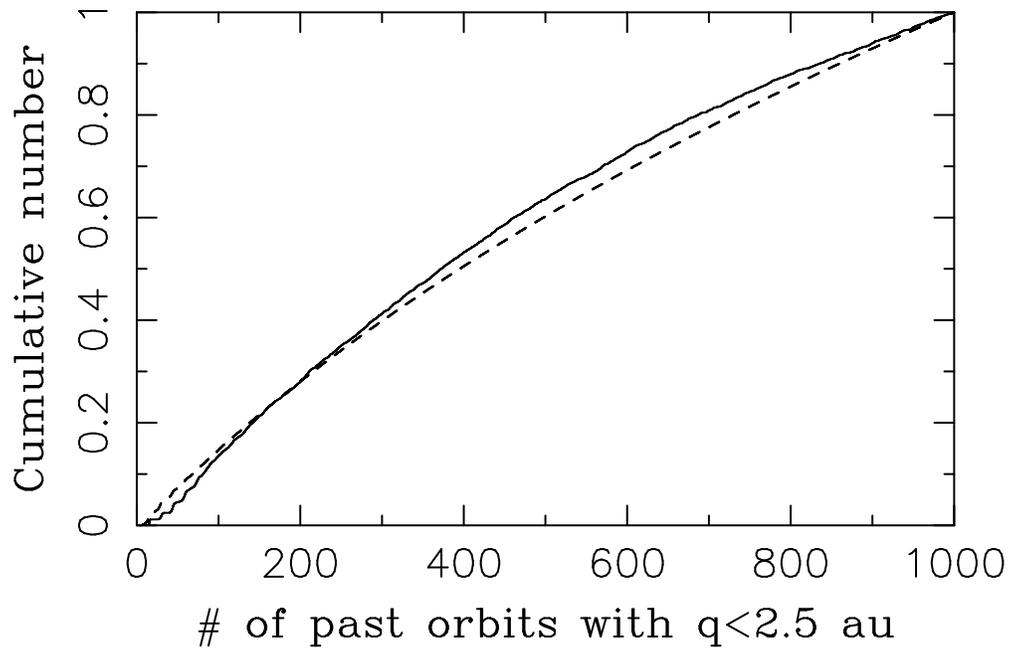}
\caption{The cumulative distribution of the number of past perihelion passages below 2.5~au. The
result for 67P is shown by the solid line. The dashed line shows the distribution for the whole 
JFC population. Here we assumed that $N_{\rm P}(2.5)=1000$.}
\label{lifet}
\end{figure}

Figure \ref{lifet} shows the cumulative distribution of the number of past perihelion passages 
with $q<2.5$ au of the whole sample. The distribution is broad and has a median of $\simeq$400
perihelion passages with $q<2.5$ au, corresponding to the median lifetime after first reaching 
$q<2.5$ au of $\simeq10^4$ yr (only about 1/3 of this time is spent on an orbit with $q<2.5$~au).\footnote{If, 
instead, we use $N_{\rm P}(2.5)=500$, which was the preferred value in N17 for the whole JFC population,  
the median number of past perihelion passages with $q<2.5$ au is found to be $\simeq200$.} This shows that 
67P probably had hundreds of perihelion passages with $q<2.5$ au in the past. Therefore, in all likelihood, 
67P is {\it not} a new comet that evolved on a JFC orbit in the past century. The probability that it 
evolved onto a JFC orbit with $q<2.5$ au for the first time in the past millennium is only $\simeq$10\%. 

Using the current mass loss of 67P, $\sim$1 meter per orbit (Paetzold et al. 2016), the above estimates 
imply that 67P lost, as an order of magnitude estimate, $\sim$400 m of surface layer due to its past 
activity. If so, the total {\it initial} volume of 67P, before 67P has become active for the first time, 
was roughly equivalent to that of a 2-km-radius sphere.

As for the collisional survival of 67P, we use the results of MR15 as a guideline. Assuming that 67P 
formed in the outer planetesimal disk below 30 au (N17) and that the disk lifetime was 400 Myr (i.e., the
{\it late} disk removal), MR15 found that a body of 67P size is expected to suffer 12-40 disruptive collisions. 
This estimate applies for the cumulative size distribution of projectiles $N(>\!\!\! D)\propto R^q$ with 
$q=-2$. Steeper (shallower) size distributions lead to a larger (smaller) number of catastrophic 
collisions.

Recent work suggests that Neptune's migration into the outer planetesimal disk and the disk dispersal 
happened early, not late (Kaib \& Sheppard 2016, Nesvorn\'y et al. 2017b, Morbidelli et al. 2018). If so, 
the lifetime of the outer disk was shorter than 400 Myr adopted in MR15, possibly much shorter. 
Assuming, for example, that the disk 
lifetime was 10 Myr, the number of catastrophic collisions obtained in MR15 would need be divided by a 
factor of 40. In addition, things depend on the dynamical state of the outer disk,
which controls the collisional probabilities and impact speeds. The planetesimal disk is expected to 
start dynamically cold (in the accretion regime) and be gradually excited by migrating Neptune and 
$\sim$1000-4000 Pluto-class objects that formed in the disk (Nesvorn\'y \& Vokrouhlick\'y 2016). For their 
nominal estimates, MR15 used the dynamical state of the disk at $t=300$~Myr from Levison et al. (2011), 
which may be a good proxy for the long term average if the planet migration/instability happened at 400 Myr. 

If, instead, the planetary migration/instability happened early, the disk remained dynamically cold during 
much of its lifetime. This may contribute by another reduction factor of at least 
2 in the number of catastrophic collisions. Also, MR15 assumed, after Brasser \& Morbidelli (2013), 
that there were $2\times10^{11}$ $D>2.3$ km bodies in the original disk, while the most recent estimates 
suggest $\simeq10^{11}$ $D>2.3$ km cometesimals (N17). Finally, the paucity of small Charon craters (Singer 
et al. 2018) indicates that the Kuiper belt may be deficient, relative to $N(>\!\!\! D)\propto R^{-2}$, in 
small projectiles (diameters $D\lesssim 1$~km). This is significant, because 67P can be collisionally 
disrupted by sub-km projectiles (MR15), and the paucity of these projectiles would imply fewer catastrophic 
collisions. 

The various factors discussed above may reduce the number of catastrophic impacts by at least $\sim 40 
\times 2 \times 2=160$. On the other hand, in MR15, the specific energy for disruption was taken from the 
strong ice case and impact speed 1 km s$^{-1}$ in Benz \& Asphaug (1999), while 67P is weaker and the 
impact speeds were lower. It is not clear what a more realistic disruption law should be and how the MR15 
results would be modified if that law is used. Jutzi et al. (2017) suggested a disruption 
law, where a porous target such as 67P is stronger, by at least a factor of $\sim$2, than the disruption
law from Benz \& Asphaug (1999). If so, this would further reduce the number of catastrophic impacts. 
In any case, taking $\sim$100 as a tentative reduction factor, 
and scaling down from the results of MR15, we find that 67P would experience only 0.1-0.4 disruptive 
collisions over 10 Myr. If so, the probability to avoid one such collisions is $\exp(-0.1)$ to $\exp(-0.4)$, 
or 0.7-0.9. Thus, in this case, the survival chances of 67P would be relatively good. A similar result
was obtained in Jutzi et al. (2017). 

Jutzi et al. (2017) assumed that the outer disk was dispersed by Neptune immediately after the 
dispersal of the protosolar nebula, and modeled the collisional evolution over the following 4.5 Gyr. Adopting 
$q=-2$, they found that the 67P-size body has a 45\% chance to avoid a catastrophic disruption, which is in 
broad agreement with the discussion above. Jutzi et al. (2017) also argued, however, that 67P would have 
suffered 14-35 {\it reshaping} impacts (for $q=-2$,  the exact number depends on the strength of 67P). If so, 
the bi-lobed shape of 67P cannot date back to the earliest stages. On the other hand, the number of reshaping 
impacts is a sensitive function of the unknown number of very small, $\sim$100 m projectiles in the Kuiper 
belt, and the paucity of small Charon craters seems to indicate that these projectiles are rare (Singer et 
al. 2018). This may imply that the number of reshaping impacts was much smaller than estimated by Jutzi et al.
(2017).

\section{Method}

Our baseline model is that 67P originally formed as a binary (see discussion in Section 5) and later, by 
perturbations caused by impacts
or dynamical effects, evolved to become a contact binary. Here we describe the methods that we used to 
estimate to likelihood of each process, planetary encounters and impacts, to end up as a contact binary.

\subsection{Planetary encounters}

N17 developed a model for the origin and dynamical evolution of JFCs. Their simulations started at the 
time of the protoplanetary gas disk dispersal. The outer planets were assumed to have an initially more 
compact configuration with Neptune at $\simeq$22 au. The outer disk of planetesimals was placed at 22-30
au. The outer extension of the disk beyond 30 au was ignored because various constraints indicate that
a great majority of planetesimals started at $<$30 au (see N17 for a discussion). The planetesimal disk
was given the mass of 15-20 $M_\oplus$, where $M_\oplus$ is the Earth mass. The disk mass is constrained by the
self-consistent simulations of the planetary migration/instability (e.g., Nesvorn\'y \& Morbidelli 2012,
Deienno et al. 2017) and by Jupiter Trojans (Nesvorn\'y et al. 2013). The size frequency distribution (SFD) of 
planetesimals was assumed to be a scaled-up version of Jupiter Trojans (Morbidelli et al. 2009a). Each 
simulation started with $10^6$ disk bodies distributed at 22-30 au with the surface density $\Sigma \propto 1/a$. 

A two-stage planetary migration/instability model was adopted from Nesvorn\'y \& Morbidelli (2012). During the 
first stage, lasting some 10-30 Myr, planets were migrated on an exponential $e$-folding timescale $\tau_1$.
The dynamical instability was assumed to happen at 10-30 Myr after the start of the simulation. During the 
instability, Neptune's orbit was modified (see discussion in N17). The integration was then continued 
with planets migrating to their present orbits on an $e$-folding timescale $\tau_2$. Eventually, all planets 
and disk bodies were evolved to $t=4.5$ Gyr after the gas disk dispersal. The integrations included 
Galactic tides and stellar encounters. 

In the last integration segment, N17 tracked bodies evolving into the inner solar system. To obtain an 
adequate statistics, bodies reaching orbits with $q<9$ au and $a<34$ au were cloned 100 times. N17 used 
different assumptions on the physical lifetime of active JFCs, including $N_{\rm p}(2.5)$ (see Section 2),
the time spent below 2.5 au, or the heliocentric distance weighted effective erosion time, where comets 
reaching very low perihelion distances were penalized. The results were compared to the known population 
of active JFCs. N17 found that different parametrizations of the physical lifetime give similar results.
Expressed in terms of $N_{\rm p}(2.5)$, the model implies that $\sim$1 km JFCs should have 300-800 perihelion 
passages below 2.5 au before becoming dormant or disrupting, while $\sim$10 km JFCs should live longer
($N_{\rm p}(2.5) \sim 3000$). 

Here we repeated two simulations from N17 and monitored {\it encounters between the disk bodies
and planets}. The two simulations had $\tau_1=10$ Myr and $\tau_2=30$ Myr (Case A) and $\tau_1=30$~Myr and 
$\tau_2=100$ Myr (Case B). This covers the interesting range of migration speeds that were inferred  
from the orbital distribution of the Kuiper belt (Nesvorn\'y 2015) and giant planet obliquities
(Vokrouhlick\'y \& Nesvorn\'y 2015; see also Bou\'e et al. 2009).   
For each encounter within $0.5 R_{{\rm H},j}$, where $R_{{\rm H},j}$ is the Hill radius 
of $j$-th planet ($j=5$ to 8 from Jupiter to Saturn; the terrestrial planets were not included), 
the planetocentric orbit of each body was recorded. We selected bodies that became 
active JFCs in the simulation (according to the criteria of N17). In the second set of simulations, 
each selected body was assumed to be a binary. The two components of each binary were given masses 
$M_1=1.4\times10^{16}$ g and $M_1=6.0\times10^{15}$ and radii $R_1=1.78$ km and $R_2=1.33$ km. This corresponds 
to the physical characteristics of the two lobes of 67P, where the radii and masses were slightly increased 
to accommodate the estimated past loss of material (Section 2).

To keep things simple, the initial eccentricities of binary orbits were set to zero and the inclinations were
selected at random (assuming the isotropic orientation of the binary orbit normal vectors). In different runs,
we varied the initial binary semimajor axis, $a_{\rm B}$, between $1<a_{\rm B}/(R_1+R_2)<1000$, or equivalently
$3.11<a_{\rm B}<3110$ km. This covers the whole range of possible initial separations. For reference, the 
heliocentric Hill radii of an object with the mass $M_1+M_2=2\times10^{16}$ g at 5 and 25 au are roughly 
1,100 and 5,600 km, respectively. 

Each binary cometesimal was evolved through each encounter recorded in the original simulation. We used 
the Bulirsch-Stoer $N$-body integrator that was adapted from the Numerical Recipes (Press et al. 1992). The 
center of mass of the binary cometesimal was first integrated backward from the time of the closest
approach to 3 $R_{{\rm H},j}$. It was then replaced by the actual binary and integrated forward through
the encounter until the planetocentric distance of the binary exceeded 3 $R_{{\rm H},j}$. The final 
binary orbit was used as the initial orbit for the next encounter and the algorithm was repeated over 
all encounters. The gravity of the Sun and other planets not having an encounter was neglected in these
integrations. The tidal evolution of binaries and precession of the binary orbit due to the non-spherical
shape of binary components was ignored as well.

We monitored collisions between binary components. If a collision occurred, the integration was stopped and the
impact speed and angle were recorded. The binary orbits that became hyperbolic during
some stage of the planetary encounter sequence were deemed to be unbound. For the surviving binaries, we 
recorded the final semimajor axis and eccentricity, which is useful to understand how much perturbation each 
binary suffered due to planetary encounters. After all integration finished, we combined individual runs 
into a statistical ensemble of evolutions that expresses the dynamical effects of planetary encounters on 
binaries. In Section 4, we use these results to discuss the possibility that the bi-lobed shape of 67P 
is a result of the collapse of 67P's parent binary triggered by planetary encounters.
 
\subsection{Impacts}

A small impact into one of the components of a binary can change the binary orbit (Petit \& Mousis 2004). 
The effect of impacts can be especially important for the small and/or loosely-bound binaries. For example, 
the two lobes of 67P, if separated by 20~km from each other, would have the orbital speed of mere 
$v_{\rm B}\simeq0.26$~m~s$^{-1}$. If a velocity change of this magnitude is delivered to one of the components, 
the binary would cease to exist. Here we investigate this process using the collision code that we previously 
developed (Morbidelli et al. 2009b, Nesvorn\'y et al. 2011). 

The code, known as {\it Boulder} (Morbidelli et al. 2009b), is a statistical particle-in-the-box 
algorithm that is capable of simulating collisional fragmentation of multiple planetesimal populations. 
It was developed along the lines of other published codes (e.g., Weidenschilling et al. 1997, Kenyon \& Bromley
2001). A full description of the {\it Boulder} code, tests, and various applications can be found in 
Morbidelli et al. (2009b), Levison et al. (2009) and Bottke et al. (2010). 

In brief, for a given impact between a projectile and a target body, the algorithm computes the specific impact 
energy $Q$, defined as the kinetic energy of the projectile divided by the total (projectile plus target) mass. 
It also computes the critical impact energy, $Q^*_{\rm D}$, defined as the energy per unit mass needed to disrupt 
and disperse 50\% of the target. For each collision, the mass of the largest remnant is computed from the scaling 
laws (e.g., Benz \& Asphaug 1999, Leinhardt \& Stewart 2009, Stewart \& Leinhardt 2009, Jutzi et al. 2017). 
The mass of the largest fragment and the slope of the power-law SFD of smaller fragments is set as function of 
$Q/Q^*_{\rm D}$ by empirical 
fits to the results of various impact simulations (e.g., Durda et al. 2004, 2007; Nesvorn\'y et al. 2006; also see 
Bottke et al. 2010).

The $Q^*_{\rm D}$ function in {\it Boulder} was assumed to split the difference between the impact simulations of 
Benz \& Asphaug (1999), who used the strong ice, and those of Leinhardt \& Stewart (2009), who used the weak ice. 
To accomplish this, we divided $Q^*_{\rm D}$ of Benz \& Asphaug by a factor, $f_Q$, where $f_Q=1$, 3 and 
10 were used in different experiments. The main input parameters of the {\it Boulder} code are the (i) initial 
SFD of the simulated populations, (ii) intrinsic collision probability, $P_i$, and (iii) mean impact speed, $v_i$. 

The binary module in the {\it Boulder} code was described in Nesvorn\'y et al. (2011). The module accounts for small, 
non-disruptive impacts on binary components, and computes the change of the binary orbit depending
on the linear momentum of the impactor. The impact velocity vectors are assumed to be randomly oriented
in the reference frame of the binary. The changes of orbital elements, $\delta a_{\rm B}$, $\delta e_{\rm B}$ and 
$\delta i_{\rm B}$, are then computed from Eqs. (7)-(9) in Nesvorn\'y et al. (2011). The binary system is assumed 
to become unbound if $a_{\rm B}$ exceeds the Hill radius, or if $e_{\rm B}>1$. The code also monitors collisions 
between components, which occur if $q_{\rm B}=a_{\rm B}(1-e_{\rm B})<R_1+R_2$.  

\section{Results}

\subsection{Planetary encounters}

We first discuss the survival of binaries during planetary encounters. Figure \ref{dsurv} shows the 
survival probability for binaries with $M_1=1.4\times10^{16}$ g and $M_2=6\times 10^{15}$ g, and different 
separations. The results in Cases A and 
B are similar. The survival probability of tight binaries with $a_{\rm B}<10$~km is $\simeq$80\%. The 
remaining $\simeq$20\% is nearly equally split between two channels of binary removal with either 
the binary components becoming unbound, or colliding to form a contact binary.

The survival probability drops with increasing separation such that for separations larger than 100 km, 
the binary survival probability is below 10\%. Most loosely bound binaries become unbound. The tightly
and loosely bound binaries can be defined by the separation at which the survival probability is 50\%.
This happens at $a_{\rm B}\simeq30$-50 km or roughly 10-17 times the sum of the component radii, 
$R_1+R_2$. The critical distance is slightly smaller in Case A than in Case B, which is related to 
the richer history of planetary encounters in Case A. The binaries that become unbound do so typically
during the early stages of evolution when they have encounters with migrating Neptune. 

Interestingly, the probability of collision between components is not a strong function of 
separation and remains at the $\sim$10\% level for the whole range of separations studied here. This is 
a combination of two opposite trends that offset each other. On one hand, for large separations, the binary 
orbit needs to reach a very large eccentricity for the two components to collide. On the other hand, 
it is easier to reach a very large eccentricity for binaries with large separations, because the 
loosely-bound binaries are more susceptible to gravitational perturbations during planetary encounters.  

Figure \ref{colvel} shows the mean collision speed between components of the collapsed binaries. The 
collision speeds are very low, 70-90 cm s$^{-1}$. For such low speeds, impacts are 
expected to result in accretion of the binary components. This may happen instantly, during a single 
head-on collision, or after a series of grazing collisions.  If the components have sufficient 
cohesion before impacts, the end result of this process should be the formation of a contact binary.
A detailed investigation of this problem is beyond the scope of this paper.  

The significance of these results for 67P depends on the number and properties of small 
binaries that emerged from the outer planetesimal disk at the time of its dispersal by 
Neptune. If we assume, for the sake of argument, that most small planetesimals were 
binaries at this stage of evolution, Figure 3 can be used to make two predictions. First, 
even under the most optimistic assumptions, the fraction of contact binaries produced 
by planetary encounters would only be $\sim$10\% (see discussion in Section 5). Second, 
binaries with the initial separations below $\sim$100 km would have good chances of survival. 
If these binaries existed at the beginning of the planetary encounter epoch, we would 
expect to have several binary comets for each contact binary comet, which is not observed. 
This implies that the small binaries must have been largely extinct at the time of the 
planetesimal disk dispersal. Indeed, we show in the following section that impacts during 
the disk stage are expected to eliminate most small binaries.

\subsection{Impacts}

Figure \ref{boulder} shows a test run where the {\it Boulder} code was used to simulate the collisional evolution
of the outer planetesimal disk. Here we used parameters similar to MR15 to be able to compare the results with MR15 
and Section 2. Specifically, we set the intrinsic collision probability $P_i=8 \times 10^{-21}$ km$^{-2}$ yr$^{-1}$, 
mean impact speed $v_i=0.4$ km s$^{-1}$ and $f_Q=1$. The initial size distribution of cometesimals was chosen to 
be similar to the present one, to test how the current distribution would be modified. Specifically, below 
the break at $D^*=100$~km, we have $N(>\!\!\!D)=c (D/10{\rm km})^{\gamma}$ with $c=6\times 10^9$ and $\gamma=-2$. 
This gives the total initial mass of 20~$M_\oplus$ and roughly $10^{11}$ cometesimals with $D>2.3$ km.

If the planetesimal disk is assumed to live for 400 Myr, as in MR15, the number of 67P-size disk bodies is
reduced by over a factor of $\simeq$10 over the disk lifetime (Figure \ref{boulder}a). In this case, as pointed 
out by MR15, the survival of 67P is unlikely. In addition, the whole size distribution changes with the final 
profile being shallower than the initial profile. The final disk mass is $<$10 $M_\oplus$, which is a problem, 
because such a small mass is 
incompatible with the existing models of the planetary migration/instability (e.g., Nesvorn\'y \& Morbidelli 2012),
where the disk mass is required to be at least 15 $M_\oplus$. Using a more massive initial disk with a steeper 
profile could help to alleviate this issue, but we were unable to find an acceptable solution with dozens of
initial SFDs that we tested. The main difficulty arises because the more massive disks grind faster and end up
with $<$10 $M_\oplus$ even if the initial mass is large. 

The problems discussed above can be resolved if the planetesimal disk was short-lived. If we adopt, for example,
a 10-Myr disk lifetime, the number of 67P-size disk cometesimals drops only by 30\% (Figure \ref{boulder}b).
The survival of 67P is likely in this case, in agreement with the discussion in Section 2. 
In addition, the disk mass is only modestly reduced from 20~$M_\oplus$ to $\simeq$19 $M_\oplus$.
Thus, a short-lived planetesimal disk may provide a more consistent framework for the early evolution of 
the solar system than the one adopted in MR15. Similar results were already reported in Jutzi et al. (2017).
They found that the probability of 67P to avoid a disruptive collision is 30-70\% (for a 
short-lived disk and $q=-2$).

Figure \ref{coll} illustrates the survival of 67P-parent binaries during the collisional evolution of the 
outer disk. Following MR15, we adopted $P_i=8 \times 10^{-21}$ km$^{-2}$ yr$^{-1}$, $v_i=0.4$ km s$^{-1}$ 
and $f_Q=1$. The disk lifetime was assumed to be 10 Myr. In this case, the surviving binary fraction
is $\simeq2\times10^{-3}$ for the tight binaries ($a_{\rm B}<10$ km) and $\sim10^{-4}$ for the loose binaries 
($a_{\rm B}=100$-1000 km). The most likely outcome of impacts is that the binary orbit becomes unbound. For
about 30\% of the tight binaries, the binary components end up colliding with each other. This 
fraction decreases to $\sim10$\% for the wide binaries. The collisional speeds between components 
of the collapsed binaries are gentle and show a trend similar to that in Figure~\ref{colvel}. 

When the disk lifetime is increased to 20 Myr, the fraction of surviving binaries drops below $10^{-5}$,
which is the resolution limit of this study (the {\it Boulder} code was set to have $10^5$ binaries for each 
initial separation). The fraction of collapsed binaries remains $\sim$10\% for the wide binaries and
$\simeq$30\% for the tight binaries. The remaining 70-90\% of binaries become unbound. 
The results do not change much when even longer disk lifetimes are considered, but in the cases with 
$>$30 Myr lifetimes, most 67P-size cometesimals become catastrophically disrupted (NR15), and the disk's 
SFD starts to diverge from the one imposed by the observational constraints (Figure \ref{boulder}).    
  
We performed several additional runs with the {\it Boulder} code to test how the results depend on various 
parameters. For example, when $v_i$ is decreased from the nominal 0.4 km s$^{-1}$ to 0.2 km s$^{-1}$, to 
simulate the dynamically cold disk conditions, we find that the fraction of tight binaries that survive 
after 10 Myr of the collisional evolution is $\simeq$0.01, about a factor of 5 higher than in the case
with $v_i=0.4$ km s$^{-1}$. This shows that the fraction of surviving binaries is a sensitive function of $v_i$. 
The fractions of unbound and collapsed binaries with $v_i=0.2$ km s$^{-1}$ are similar to and follow the 
same trends as those obtained with $v_i=0.4$ km s$^{-1}$ (e.g., 30\% of tight binaries collapse into 
contact binaries). 

\subsection{Kozai cycles} 
  
The Kozai dynamics of a binary orbit arises due to the gravitational potential of the Sun (see Naoz et al. 
2016 for a review). In the simplest quadrupole approximation of the solar gravity field, the quantity 
$(1-e_{\rm B}^2)^{1/2} \cos i_{\rm B}$ is conserved and the problem is integrable. For a trajectory starting 
with $e_{\rm B}=0$ and $i_{{\rm B},0}$, the maximum eccentricity that can be reached during the Kozai cycles 
is $e_{\rm B,max}=(1-5/3\cos^2 i_{{\rm B},0})^{1/2}$. For the the two components to collide, $q_{\rm B}<R_1+R_2$.
This defines a critical value, $i^*_{\rm B}$, where $\cos^2 i^*_{\rm B} = 3/5(1-(1-(R_1+R_2)/a_{\rm B})^2)$.
If the initial inclination, $i_{{\rm B},0}$, is larger than $i^*_{\rm B}$, the two components will collide. If 
$i_{{\rm B},0}<i^*_{\rm B}$, on the other hand, $q_{\rm B}$ will not drop below $R_1+R_2$ and the binary
system will survive. Assuming an isotropic initial distribution of $i_{{\rm B},0}$, the survival probability 
as a function of separation is shown in Figure \ref{kozai}. 

In addition, for the Kozai cycles to be effective, the two binary components must be roughly spherical 
and/or the binary separation must be large. If not, the gravitational potential from $J_2$ of the 
binary components will prevail over the solar gravity, resulting in a simple precession of the binary orbit
pole about the heliocentric orbit pole. The critical semimajor axis is $a^*_{\rm B}=(2 \mu J_2 R^2 a_{\rm h}^3)^{1/5}$, 
where $\mu$ is the binary-to-Sun mass ratio, and $a_{\rm h}$ is the semimajor axis of the heliocentric orbit 
(e.g., Mignard 1982). The $J_2 R^2$ term is the measure of non-sphericity of the binary components. For a homogeneous 
ellipsoid with axes $a>b>c$, $J_2 R^2 = (a^2+b^2-2c^2)/10$.  Summing up the contributions from the 
measured shapes of the two lobes of 67P we have $J_2 R^2 \simeq 0.67$ km$^2$. In Figure \ref{kozai}, 
we plot $a^*_{\rm B}=260$ km corresponding to $J_2 R^2 = 0.67$ km$^2$, $\mu=10^{-17}$ and $a_{\rm H}=25$~au.

It is apparent from Figure \ref{kozai} that the survival of a 67P-parent binary is likely for all initial 
separations. For small separations, the fast apsidal precession due to $J_2$ renders the Kozai cycles 
ineffective. At large separations, the collision orbits do not represent an important volume in space of 
initial conditions, because the orbital eccentricity must become very large for the collision to occur. 
This happens only if the initial inclination of the binary orbit is very close to $90^\circ$ (relative to 
the heliocentric orbit). We therefore conclude that the Kozai dynamics should play a only minor role.
 
\section{Discussion and Conclusions}

We postulated that 67P formed as a binary and evaluated the probability that the binary orbit was destabilized 
by collisional and dynamical processes. Whether 67P actually formed as a binary is unclear. On one hand, binaries 
are common among the 100-km-class CCs at 40-50 au. The formation of CC binaries 
is thought to be related to the accretion processes in the early solar system (e.g., Goldreich et al. 2002,
Nesvorn\'y et al. 2010). The present-day fraction of binaries in the CC population is estimated to be 30-100\% (e.g., 
Noll et al. 2008, Fraser et al. 2017). The CC binaries survived to the present day because the 40-50 au region 
presumably received only modest perturbations from the collisional and dynamical processes. 

On the other hand, 67P formed significantly closer to the Sun, probably in the 20-30 au disk, and is much 
smaller than the observed CC binaries. To establish whether 67P formed as a primordial binary would 
therefore need to understand how the accretion processes scale with the heliocentric distance and size.
For example, planetesimals may have formed by the streaming instability followed by gravitational collapse
(Youdin \& Goodman 2005, Johansen \& Youdin 2007). If so, it would probably be reasonable to assume that 
their formation at 20-30 au and 40-50 au followed the same suit, because the streaming instability is not expected
to have a strong dependence on the heliocentric distance.

Also, Simon et al. (2017) and others showed that the streaming instability is capable of forming 
planetesimals of different sizes with the expected SFD scaling that is similar to the observed SFD of the 
Kuiper belt objects and Jupiter Trojans below 100 km. This may suggest that the formation of 67P-size 
cometesimals was just a scaled-down version of the formation of 100-km CCs (note that the existing 
streaming instability simulations do not have the required resolution to explicitly demonstrate the 
formation of km-size cometesimals). These arguments may provide some justification to our assumption 
that 67P formed as a binary. In contrast, the high-resolution simulations of the streaming 
instability show that small clumps of pebbles can be easily dispersed by turbulent diffusion (Klahr \& 
Schreiber 2016). If so, the formation of the 67P-size cometesimals by the streaming instability may be 
inefficient. Other accretion models, including the hierarchical coagulation by two-body 
collisions, may have different implications (see Youdin \& Kenyon 2013).

The expected fraction of contact binaries among JFCs is the product of: (1) the fraction
of cometesimals that formed as binaries, $f_{\rm binary}$, and (2) the fraction of binaries that 
collapsed to become contact binaries, $f_{\rm contact}$. The main contribution of this work was to estimate 
$f_{\rm contact}$ for the 67P-class comets. We found that small impacts during the collisional evolution 
of the outer disk at 20-30 au give $f_{\rm contact}=10$-30\% (for the range of the initial binary 
separations and outer disk lifetimes considered here). The disk lifetimes $\lesssim$10 Myr are required for 
67P to avoid a catastrophic disruption (MR15 and Section 2). Thus, for example, if $f_{\rm binary}\sim0.5$, 
the expected fraction of 67P-like contact binaries among JFCs would be $\sim$5-15\%. 

While the processes described here could potentially explain the bi-lobed shape of 67P, they are probably 
not efficient enough to explain the shapes of comets in general. Six comets were imaged by spacecrafts 
and have good shape models: 1P/Halley, 9P/Tempel 1, 19P/Borrely, 67P, 81P/Wild 2 and 103P/Hartley 2. 
Of these, 67P stands as the one with the most bi-lobed shape. Halley, Borrelly and Hartley 2 are also 
bi-lobed but less clearly so than 67P. Wild 2 and Tempel 1 are more rounded. These observations therefore 
indicate that 4 out of 6, or roughly 67\% of comets appear to be bi-lobed, which is a much larger fraction 
than the one expected from the statistics of collapsed binaries (see above). This suggests that other,
more efficient mechanism must be at play (e.g., Jutzi \& Benz 2017, Schwartz et al. 2018).   

\begin{figure}
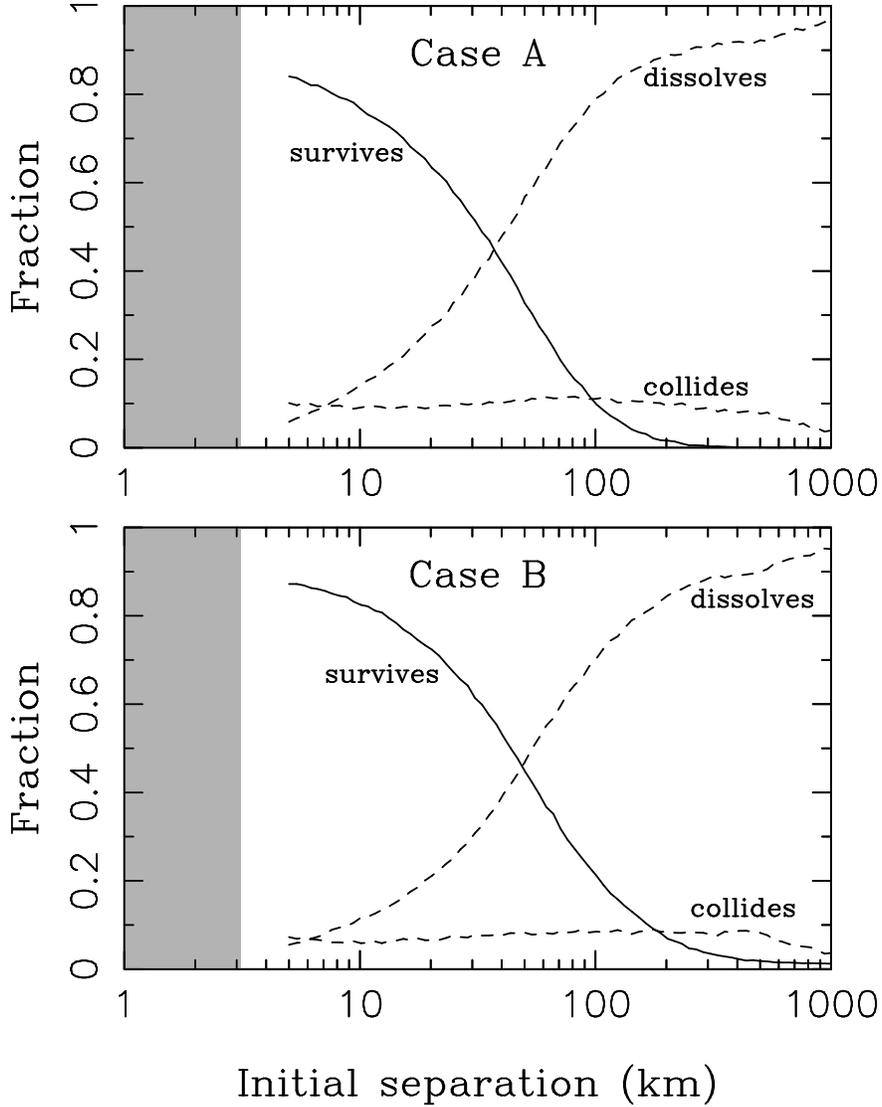

\epsscale{0.7}
\plotone{fig3a.eps}\\[3.mm]
\plotone{fig3b.eps}
\caption{The dynamical survival of 67P-parent binaries during planetary encounters. In the two migration 
cases A and B, the solid line shows the probability that a binary with given separation 
survives the whole sequence of planetary encounters. The dashed lines denote the probability that
the binary becomes unbound or collapses into a contact binary. The latter outcome happens in $\simeq$10\% 
of cases. The gray box shows where the two components are in contact. We did not investigate binaries 
with separations below 5 km, where dynamics is strongly influenced by the neglected $J_2$ term.}
\label{dsurv}
\end{figure}

\begin{figure}
\epsscale{0.8}
\plotone{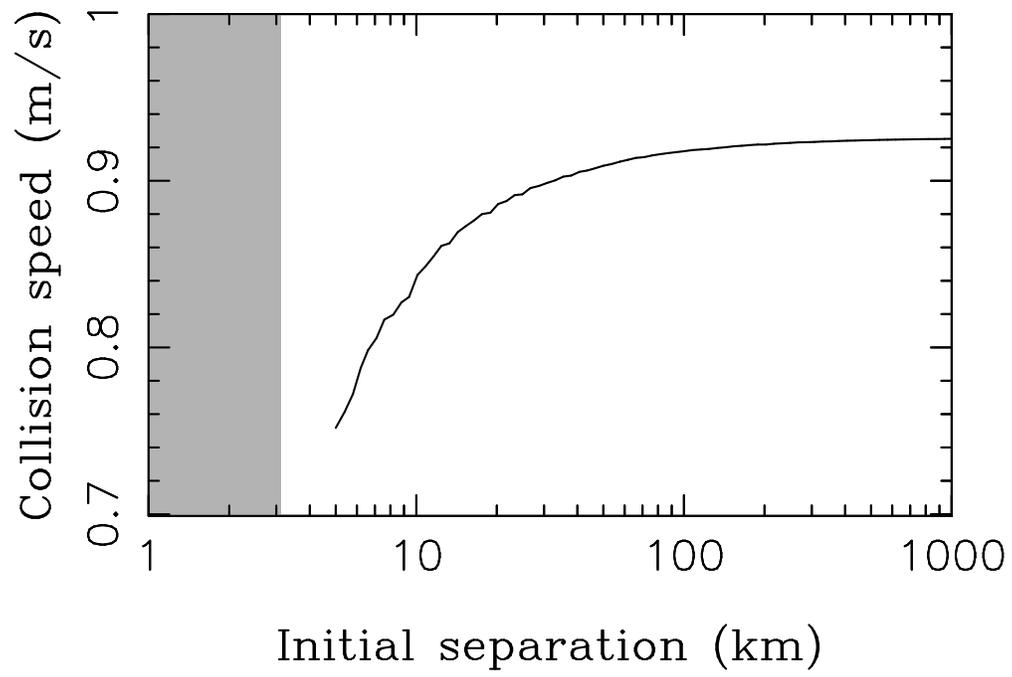}
\caption{The mean collision speed of the 67P-parent binaries whose components ended colliding with each 
other.}
\label{colvel}
\end{figure}

\begin{figure}
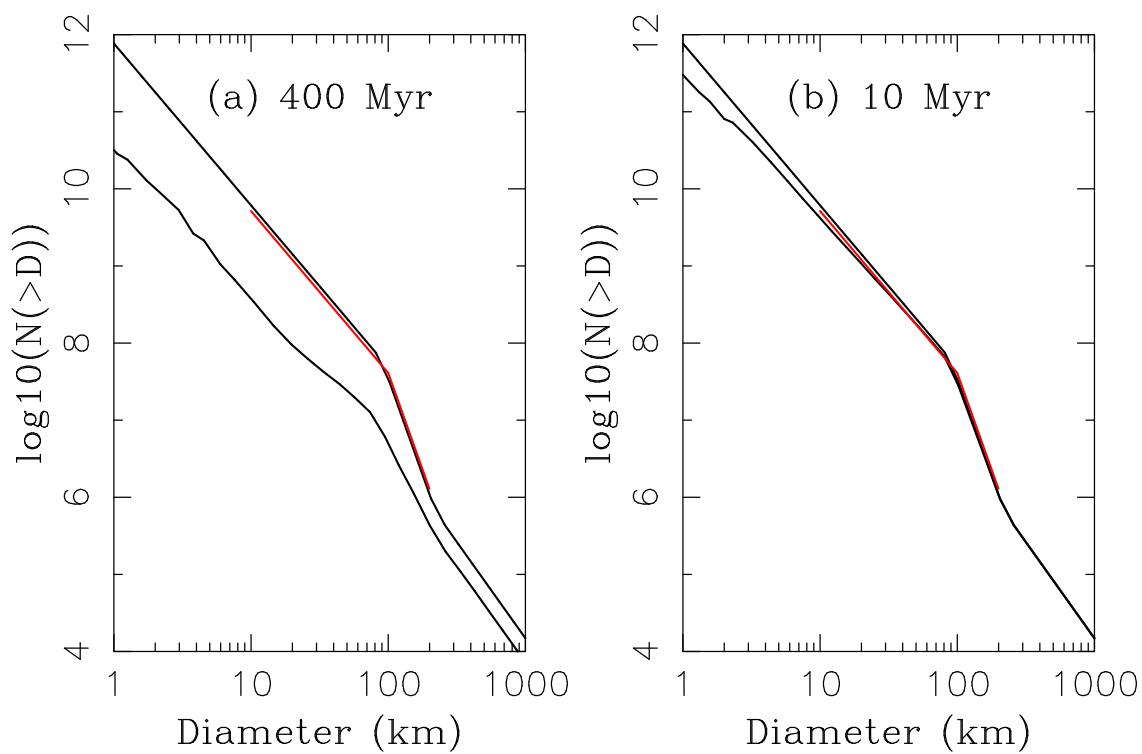

\epsscale{0.45}
\plotone{fig5a.eps}
\plotone{fig5b.eps}
\caption{The collisional evolution of the outer planetesimal disk. In panel (a), we assumed that the 
disk is long-lived (400 Myr). The initial and final size distributions are shown by the upper and lower black 
lines, respectively. The red line is the target distribution constrained by Jupiter Trojans and 
planetary migration/instability calculations. Panel (b) shows the same for a short-lived disk (10 Myr).}
\label{boulder}
\end{figure}

\begin{figure}
\epsscale{0.8}
\plotone{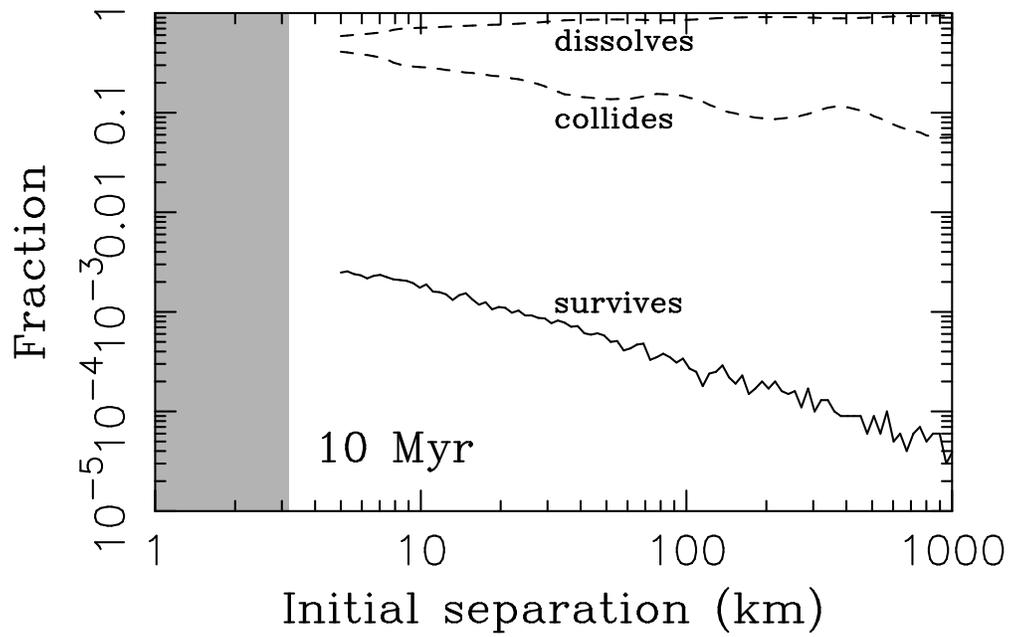}
\caption{The survival of 67P-parent binaries during the collisional evolution of the outer planetesimal disk.
Here we assumed that the outer disk lasts 10 Myr before it is dispersed by Neptune. The solid line
shows the surviving binary fraction. The dashed lines denote the fraction of binaries collapsing into
contact binaries and those becoming unbound.}
\label{coll}
\end{figure}

\begin{figure}
\epsscale{0.8}
\plotone{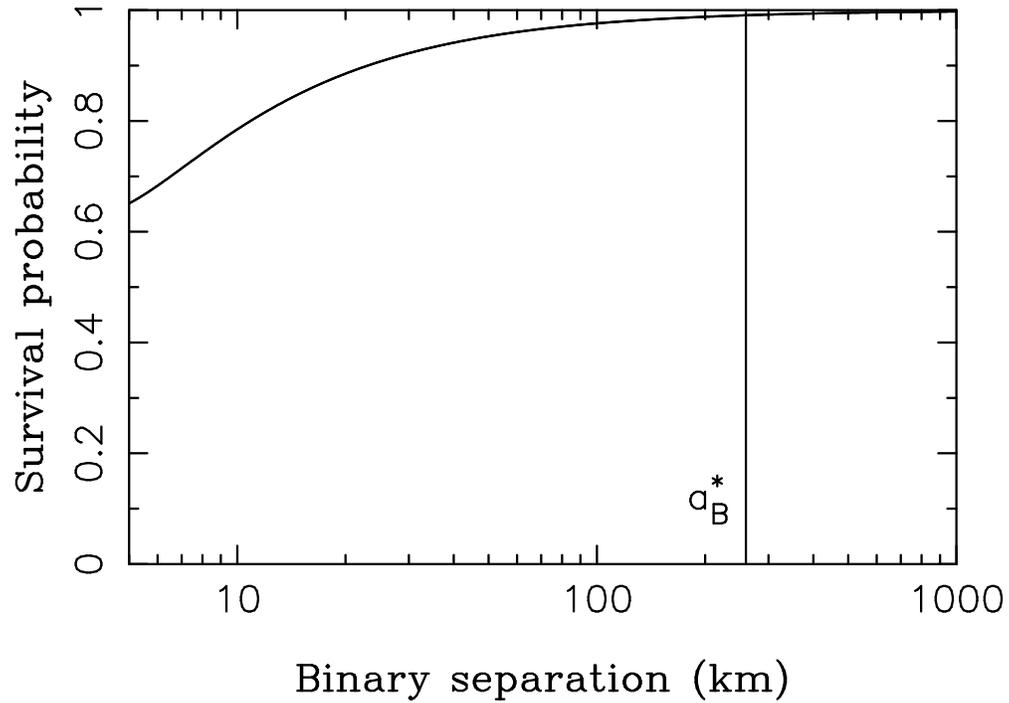}
\caption{The survival of 67P-parent binaries against Kozai-induced collisions 
between components. The initial orientations of the binary orbits were assumed to be random. The vertical 
solid line shows the transition between the $J_2$-dominated dynamics for small separations to the 
Kozai-dominated dynamics for large separations. See Section 4.3 for the parameter values adopted in 
this plot.}
\label{kozai}
\end{figure}

Only a very small fraction ($<2\times10^{-3}$ for $v_i=0.4$ km s$^{-1}$ and $\gtrsim$10-Myr lifetime) 
of 67P-parent binaries survive the collisional evolution of the outer planetesimal disk. The surviving binaries 
undergo gravitational perturbations during planetary encounters and are further reduced in number. 
The survival during planetary encounters depends on the initial binary separation: most tight binaries with 
$a_{\rm B}<30$-50 km survive, while most $a_{\rm B}>30$-50 km are dissolved (Figure \ref{dsurv}). The probability 
to become a contact binary is $\sim$10\% during this stage. Given that the number of binaries was severely 
reduced during the previous stage of the collisional evolution, the relevance of planetary encounters for 
the contact binary formation must be relatively minor. 

In addition, we find that it is very unlikely that 67P is a new comet that evolved on a JFC orbit in 
the past century. We estimate that the probability that it evolved onto an orbit with $q<2.5$ au in 
the past 1000 yr is only 10\%.

\acknowledgements
This work was supported by funding for the Rosetta-Alice project from NASA via Jet Propulsion Laboratory 
contract 1336850 to the Southwest Research Institute. We thank A. Morbidelli for a helpful referee report.

\clearpage

\clearpage

\clearpage

\clearpage

\clearpage

\clearpage

\clearpage


\begin{thebibliography}{}

\bibitem[Benz \& Asphaug(1999)]{1999Icar..142....5B} Benz, W., \& Asphaug, E.\ 1999, Icarus, 142, 5 

\bibitem[Blum et al.(2017)]{2017MNRAS.469S.755B} Blum, J., Gundlach, B., Krause, M., et al.\ 2017, \mnras, 469, S755 

\bibitem[Bottke et al.(2010)]{2010AJ....139..994B} Bottke, W.~F., Nesvorn{\'y}, D., Vokrouhlick{\'y}, D., \& Morbidelli, A.\ 2010, \aj, 139, 994 

\bibitem[Bou{\'e} et al.(2009)]{2009ApJ...702L..19B} Bou{\'e}, G., Laskar, J., \& Kuchynka, P.\ 2009, \apjl, 702, L19 

\bibitem[Brasser \& Morbidelli(2013)]{2013Icar..225...40B} Brasser, R., \& Morbidelli, A.\ 2013, Icarus, 225, 40 

\bibitem[Davidsson et al.(2016)]{2016A&A...592A..63D} Davidsson, B.~J.~R., Sierks, H., G{\"u}ttler, C., et al.\ 2016, \aap, 592, A63 

\bibitem[Deienno et al.(2017)]{2017AJ....153..153D} Deienno, R., Morbidelli, A., Gomes, R.~S., \& Nesvorn{\'y}, D.\ 2017, \aj, 153, 153 

\bibitem[Durda et al.(2004)]{2004Icar..167..382D} Durda, D.~D., Bottke, W.~F., Enke, B.~L., et al.\ 2004, Icarus, 167, 382 

\bibitem[Durda et al.(2007)]{2007Icar..186..498D} Durda, D.~D., Bottke, W.~F., Nesvorn{\'y}, D., et al.\ 2007, Icarus, 186, 498

\bibitem[Fraser et al.(2017)]{2017NatAs...1E..88F} Fraser, W.~C., Bannister, M.~T., Pike, R.~E., et al.\ 2017, Nature Astronomy, 1, 0088 

\bibitem[Goldreich et al.(2002)]{2002Natur.420..643G} Goldreich, P., Lithwick, Y., \& Sari, R.\ 2002, \nat, 420, 643 

\bibitem[Guzzo \& Lega(2017)]{2017MNRAS.469S.321G} Guzzo, M., \& Lega, E.\ 2017, \mnras, 469, S321 

\bibitem[Johansen et al.(2012)]{2012A&A...537A.125J} Johansen, A., Youdin, A.~N., \& Lithwick, Y.\ 2012, \aap, 537, A125 

\bibitem[Johansen et al.(2009)]{2009ApJ...704L..75J} Johansen, A., Youdin, A., \& Mac Low, M.-M.\ 2009, \apjl, 704, L75 

\bibitem[Jorda et al.(2016)]{2016Icar..277..257J} Jorda, L., Gaskell, R., Capanna, C., et al.\ 2016, Icarus, 277, 257 

\bibitem[Jutzi \& Benz(2017)]{2017A&A...597A..62J} Jutzi, M., \& Benz, W.\ 2017, \aap, 597, A62 

\bibitem[Jutzi et al.(2017)]{2017A&A...597A..61J} Jutzi, M., Benz, W., Toliou, A., Morbidelli, A., \& Brasser, R.\ 
2017, \aap, 597, A61 

\bibitem[Kaib \& Sheppard(2016)]{2016AJ....152..133K} Kaib, N.~A., \& Sheppard, S.~S.\ 2016, \aj, 152, 133 

\bibitem[Kenyon \& Bromley(2001)]{2001AJ....121..538K} Kenyon, S.~J., \& Bromley, B.~C.\ 2001, \aj, 121, 538 

\bibitem[Klahr \& Schreiber(2016)]{2016IAUS..318....1K} Klahr, H., \& Schreiber, A.\ 2016, Asteroids: New Observations, 
New Models, 318, 1 

\bibitem[Leinhardt \& Stewart(2009)]{2009Icar..199..542L} Leinhardt, Z.~M., \& Stewart, S.~T.\ 2009, Icarus, 199, 542 

\bibitem[Levison \& Duncan(1997)]{1997Icar..127...13L} Levison, H.~F., \& Duncan, M.~J.\ 1997, Icarus, 127, 13 

\bibitem[Levison et al.(2009)]{2009Natur.460..364L} Levison, H.~F., Bottke, W.~F., Gounelle, M., et al.\ 2009, \nat, 460, 364 

\bibitem[Levison et al.(2011)]{2011AJ....142..152L} Levison, H.~F., Morbidelli, A., Tsiganis, K., Nesvorn{\'y}, D., \& Gomes, R.\ 2011, \aj, 142, 152 

\bibitem[Maquet(2015)]{2015A&A...579A..78M} Maquet, L.\ 2015, \aap, 579, A78 

\bibitem[Mignard(1982)]{1982Icar...49..347M} Mignard, F.\ 1982, Icarus, 49, 347 

\bibitem[Morbidelli \& Rickman(2015)]{2015A&A...583A..43M} Morbidelli, A., \& Rickman, H.\ 2015 (MR15), \aap, 583, A43 

\bibitem[Morbidelli et al.(2009)]{2009Icar..204..558M} Morbidelli, A., Bottke, W.~F., Nesvorn{\'y}, D., \& Levison, H.~F.\ 2009a, Icarus, 204, 558 

\bibitem[Morbidelli et al.(2009)]{2009Icar..202..310M} Morbidelli, A., Levison, H.~F., Bottke, W.~F., Dones, L., \& Nesvorn{\'y}, D.\ 2009b, Icarus, 202, 310 

\bibitem[Morbidelli]{2015A&A...583A..43N} Morbidelli, A., Nesvorn\'y, D., et al., Icarus, in press 

\bibitem[Naoz(2016)]{2016ARA&A..54..441N} Naoz, S.\ 2016, \araa, 54, 441 

\bibitem[Nesvorn{\'y}(2015)]{2015AJ....150...73N} Nesvorn{\'y}, D.\ 2015, \aj, 150, 73 

\bibitem[Nesvorn{\'y} \& Morbidelli(2012)]{2012AJ....144..117N} Nesvorn{\'y}, D., \& Morbidelli, A.\ 2012, \aj, 144, 117 

\bibitem[Nesvorn{\'y} \& Vokrouhlick{\'y}(2016)]{2016ApJ...825...94N} Nesvorn{\'y}, D., \& Vokrouhlick{\'y}, D.\ 2016, \apj, 825, 94 

\bibitem[Nesvorn{\'y} et al.(2006)]{2006Icar..183..296N} Nesvorn{\'y}, D., Enke, B.~L., Bottke, W.~F., et al.\ 2006, Icarus, 183, 296 

\bibitem[Nesvorn{\'y} et al.(2010)]{2010AJ....140..785N} Nesvorn{\'y}, D., Youdin, A.~N., \& Richardson, D.~C.\ 2010, \aj, 140, 785

\bibitem[Nesvorn{\'y} et al.(2011)]{2011AJ....141..159N} Nesvorn{\'y}, D., Vokrouhlick{\'y}, D., Bottke, W.~F., Noll, K., \& Levison, H.~F.\ 2011, \aj, 141, 159 

\bibitem[Nesvorn{\'y} et al.(2013)]{2013ApJ...768...45N} Nesvorn{\'y}, D., Vokrouhlick{\'y}, D., \& Morbidelli, A.\ 2013, \apj, 768, 45 

\bibitem[Nesvorn{\'y} et al.(2017)]{2017ApJ...845...27N} Nesvorn{\'y}, D., Vokrouhlick{\'y}, D., Dones, L., et al.\ 2017 
(N17), \apj, 845, 27 

\bibitem[Nesvorn{\'y} et al.(2017)]{2017AJ....153..103N} Nesvorn{\'y}, D., Roig, F., \& Bottke, W.~F.\ 2017, \aj, 153, 103 

\bibitem[Noll et al.(2008)]{2008ssbn.book..345N} Noll, K.~S., Grundy, W.~M., Chiang, E.~I., Margot, J.-L., \& Kern, S.~D.\ 2008, The Solar System Beyond Neptune, 345 

\bibitem[Paetzold et al.(2016)]{2016DPS....4811627P} Paetzold, M., Andert, T., Hahn, M., et al.\ 2016, AAS/Division for Planetary Sciences Meeting Abstracts, 48, 116.27 

\bibitem[Parker \& Kavelaars(2010)]{2010ApJ...722L.204P} Parker, A.~H., \& Kavelaars, J.~J.\ 2010, \apjl, 722, L204 

\bibitem[Press et al.(1992)]{1992nrfa.book.....P} Press, W.~H., Teukolsky, S.~A., Vetterling, W.~T., \& Flannery, B.~P.\ 1992, Cambridge: University Press, |c1992, 2nd ed.,  

\bibitem[Rickman et al.(2015)]{2015A&A...583A..44R} Rickman, H., Marchi, S., A'Hearn, M.~F., et al.\ 2015, \aap, 583, A44 

\bibitem[Schwartz et al.(2018)]{2018NatAs.tmp....1S} Schwartz, S.~R., Michel, P., Jutzi, M., et al.\ 2018, Nature Astronomy,
in press  

\bibitem[Simon et al.(2017)]{2017ApJ...847L..12S} Simon, J.~B., Armitage, P.~J., Youdin, A.~N., \& Li, R.\ 2017, \apjl, 847, L12

\bibitem[Singer et al.(2016)]{2016DPS....4821312S} Singer, K.~N., McKinnon, W.~B., Gladman, B., et al.\ 2018, Science, under review 

\bibitem[Stewart \& Leinhardt(2009)]{2009ApJ...691L.133S} Stewart, S.~T., \& Leinhardt, Z.~M.\ 2009, \apjl, 691, L133 

\bibitem[Vokrouhlick{\'y} \& Nesvorn{\'y}(2015)]{2015ApJ...806..143V} Vokrouhlick{\'y}, D., \& Nesvorn{\'y}, D.\ 2015, \apj, 806, 143 

\bibitem[Weidenschilling et al.(1997)]{1997Icar..128..429W} Weidenschilling, S.~J., Spaute, D., Davis, D.~R., Marzari, F., \& Ohtsuki, K.\ 1997, Icarus, 128, 429 
 
\bibitem[Youdin \& Kenyon(2013)]{2013pss3.book....1Y} Youdin, A.~N., \& Kenyon, S.~J.\ 2013, Planets, Stars and Stellar Systems.~Volume 3: Solar and Stellar Planetary Systems, 1 

\bibitem[Youdin \& Johansen(2007)]{2007ApJ...662..613Y} Youdin, A., \& Johansen, A.\ 2007, \apj, 662, 613 

\bibitem[Youdin \& Goodman(2005)]{2005ApJ...620..459Y} Youdin, A.~N., \& Goodman, J.\ 2005, \apj, 620, 459 
 
\end{thebibliography}
\end{document}